\documentclass[11pt]{article}
\usepackage{moriond,epsfig}

\def\be{\begin{equation}}
\def\ee{\end{equation}}
\def\bea{\begin{eqnarray}}
\def\eea{\end{eqnarray}}

\begin{document}
\vspace*{4cm}
\title{UNRAVELING VIOLATIONS OF PARTON-HADRON DUALITY IN $ep$ SCATTERING}

\author{ S. LIUTI }

\address{Department of Physics, University of Virginia, 382 McCormick Road,\\
Charlottesville, VA 22904-4714, USA}

\maketitle\abstracts{Recent studies of the $Q^2$ dependence of $ep$ scattering
in the large $x$ region and in the $Q^2$ range: $1 \leq Q^2 \leq 30$ 
Gev$^2$, confirm the validity of the phenomenon of quark-hadron duality -
the similarity of the deep inelastic (parton) and resonance (hadron) 
spectra - for values of the invariant mass, $W^2 \geq 2.4$ GeV$^2$. 
At lower values of $W^2$, duality is found to be significantly violated by an
amount that cannot be parametrized in terms of the first few terms of 
a series of power corrections. We present a dynamical model that explains the
$Q^2$ dependence of the data: at low $W^2$, non-partonic components given by
color neutral clusters dominate the cross section and the $Q^2$ dependence is
governed by their mass spectrum, predicted within the preconfiment property of QCD; 
at large $W^2$ the structure function is determined by a convolution of the 
cluster mass spectrum with the parton distributions.}

\section{Introduction}

Deep inelastic scattering (DIS) experiments provide a wealth of 
information on the structure of hadrons, from a detailed understanding of their
partonic structure, to precision measurements of $\alpha_S$.  
In order to reveal the partonic
structure of hadrons, the four-momentum transfer squared,
$Q^2$, must be large, namely well above 1 GeV$^2$.    
%Although hard scattering processes are envisaged as  
%nearly head-on and almost instantaneous collisions between partons inside 
%the hadron and their probes,
%a rich dynamics ensues, as a result of gluon and $q \bar{q}$ pair production  
%by the struck parton. These processes are described within perturbative QCD (pQCD).
As $Q^2$ is lowered partonic components become less likely
to be observed, and a transition to a regime eventually 
dominated by {\em non-partonic} degrees of freedom,
and therefore outside the range of applicability of pQCD, 
is expected to occur.  
The onset of this regime has been observed \cite{Breit} in a large number of experiments
on $ep$ scattering in the low Bjorken $x$ and large invariant mass, $W^2$, region, where 
it is a well accepted fact that non-partonic components should 
replace the partonic structure at low $Q^2$. 
The data indeed support a sharp transition between the partonic and non-partonic
regimes at $Q^2 \approx 1$ GeV$^2$, for $ 10^{-5} \leq x \leq 10^{-3}$
.
At large $x$ the proton structure function is clearly dominated by non-partonic 
components -- the nucleon resonances  -- up to
relatively large $Q^2$ ($Q^2 \leq 20$ GeV$^2$).
An equivalence was singled out, however, between 
the average of the resonance spectrum, written in terms of Mellin moments, 
and the DIS structure function measured at larger $Q^2$. 
The moments in the two regimes were found to differ by 
perturbative corrections and relatively small power corrections \cite{DGP}.
It was conjectured that a form of {\em duality} resulted from the cancellation of higher order 
terms in the twist expansion that would otherwise be expected to 
dominate the cross section at $x \rightarrow 1$, or as more exclusive states are produced.
%The data were shown to be fitted by allowing for a residual, small, ${\cal O}(1/Q^2)$ term. 
This view has been since considered the most natural one. 
%Nevertheless a full dynamical explanation for such a cancellation is still lacking.
%This is not surprising -- the phenomenon of parton-hadron duality is
%deeply connected with the intensively investigated 
%conceptual duality on which QCD itself is founded, and by which
%asymptotic freedom coexists with permanently confined quarks and gluons.
From a slightly different perspective, a series of recent papers 
has been devoted to {\em local} quark-hadron duality and 
its violations in semi-leptonic decays, and $\tau$ decays (for a review see \cite{BIGI}). 
In particular, it was shown in \cite{SHI} that local duality violations 
can be traced to the asymptotic nature of the operator product expansion (OPE), 
namely, to the behavior of operators of both higher dimension and higher twist. 
This has lead the way to specific models based on instantons, and on large 
$N_c$ QCD in $(1+1)$ dimensions. 
%and on quark models \cite{Yao}, 
%addressed specifically at gauging the violations of local 
%duality and their possible impact on the experimental extraction of CKM matrix elements. 
%The framework developed in \cite{SHI} parallels ideas initially put forth for DIS. However, 
%practical results are based on the Small Velocity approximation, which is specific to heavy 
%quarks interactions. 
Approaches consistent with \cite{BIGI,SHI} could shed some light on duality in DIS, 
where quark-hadron duality has been explored so far within the context of 
quark models \cite{Wal}, thus avoiding 
the basic questions related to the dynamical nature of QCD. 
%provided practical avenues are devised. 
%This would represent an alternative approach to quark model, the only  
%have been explored so far \cite{Wal}, thus avoiding the basic questions
%related to the gauge nature of QCD. 
%%% MY PROPOSAL 
In this contribution, we discuss a possible avenue: Our starting point
is similar to \cite{BIGI,SHI}, in that the background of our model is the OPE within which 
we pursue a connection between the resonance region and the higher twist operators.
% aimed at a better taking into account the gauge nature of QCD.
%, thus avoiding the basic questions
%related to the gauge nature of QCD. 
Crucial for the construction of our model is an accurate analysis of 
recent data \cite{ioana1} conducted in \cite{SIMO1}. 
In the following sections we summarize the results found in \cite{SIMO1},   
we introduce our model, and we present some initial results. 

\section{Perturbative QCD Analysis of Parton-Hadron Duality}

In \cite{SIMO1} the $Q^2$ dependence in the resonance region was extracted 
by first considering the average of the resonances over $\xi=2x/(1+\sqrt{1+4M^2x^2/Q^2})$, which 
properly takes into account Target Mass Corrections (TMC). The averaging procedure, described in detail
in \cite{ioana1}, yields a smooth curve in $\xi$ which fits the resonance data 
with a $\chi^2/d.o.f.$ between 0.8 and 1.1. It represents an alternative analysis to the ones 
using moments.  
The fit was performed in bins of $W^2$, centered at: $W_R^2=$ 1.6, 2.3, 2.8, 3.4 GeV$^2$, respectively. 
Our subsequent study was aimed at establishing whether it is possible to define 
a breakpoint where pQCD no longer applies
and a transition occurs, similar to what observed 
in the low $x$ regime \cite{Breit}.
%We performed a detailed study of both 
%the logarithmic and  power corrections in order to ascertain whether the apparent 
%weak $Q^2$ dependence of the data in the low $W^2$ region reported in \cite{ioana1} 
%is coincidental, an artifact of the particular 
%region under study, or a cancellation of HT
%terms, possibly understandable within parton-hadron duality models.     
The analysis involved a number of steps similar to recent 
extractions of power corrections from inclusive data 
\cite{EXP_HT,AKandco,twistpaper}, namely the form 
\begin{equation}
F_2^{exp}(x,Q^2) = F_2^{pQCD+TMC}(x,Q^2) + \frac{H(x,Q^2)}{Q^2} +{\cal O}(1/Q^4),   
\label{t-expansion}
\end{equation}
was adopted, where $F_2^{pQCD+TMC}(x,Q^2)$ is the twist-2 contribution, including  
the kinematical TMC; the other terms in the 
formula are the dynamical power corrections, formally arising 
from higher order terms in the twist expansion. 
Both $F_2^{pQCD+TMC}$ and $H$ were extracted from the data
at large $x$.
The shape of the initial NS PDFs 
was found to be well constrained  at variance with
the singlet and gluon distributions at low $Q^2$, 
whose shape is strongly correlated with the value of $\alpha_S$.
NNLO corrections were not included.
These would introduce further theoretical uncertainties. In fact,  
the question of whether  
they can ``mimick'' the contributions of higher twists, including 
the uncertainties due to the well known scale/scheme dependence of 
calculations, within the current precision of data is still a subject of intense 
investigations \cite{AKandco}.
Large $x$ resummation was performed directly in $x$ 
space by replacing the $Q^2$ scale 
with a $z$-dependent one, $\widetilde{W}^2 = Q^2(1-z)/z$ \cite{Rob_pap}. 
It was found that over the range 
$0.45 \leq x \leq 0.85$, large $x$ resummation, and 
TMC improve the agreement with the data.
We parametrized the remaining discrepancies through 
the Higher Twist (HT) coefficient 
$H(x,Q^2) = F_2^{pQCD+TMC}(x,Q^2) C_{HT}(x)$. 
In Fig. 1a we show $C_{HT}$, extracted from:
DIS data with $W^2 \ge 4$ GeV$^2$, from the resonance region ($W^2 < 4$ GeV$^2$),  
and over the entire range of $W^2$.     
The figure also shows the range of extractions previous to the current one 
\cite{EXP_HT}.
While the large $W^2$ data track a curve that is 
consistent with the $1/W^2$ behavior expected from most models, the low $W^2$ data yield a much smaller 
value for $C_{HT}$ and they show a bend-over of
the slope vs. $x$.
This surprising effect is not a consequence 
of the interplay of higher order corrections
and the HT terms, but just  
of the extension of our detailed pQCD analysis to 
the large $x$, low $W^2$  kinematical region. 
In other words, we unraveled a $Q^2$ dependence that seems to deviate 
from the common wisdom developed since the pioneering analysis of \cite{DGP}, or,
in the language of \cite{SHI}, we observe a violation of {\em global} duality.  
Similar results, expressed through a $Q^2$ dependent HT coefficient, were also obtained
in \cite{SzcUle}.

\section{Large $N_c$ Model for Initial Parton Evolution}

We propose a simple dynamical model for the structure function
in the low $W^2$ ($W^2 \leq 4$ GeV$^2$) and low $Q^2$ ($Q^2 < 10 $ GeV$^2$) regime, where non-partonic 
configurations are expected to be dominant. 
The DIS cross section for $ep$ scattering is proportional to the structure functions
$F_{1(2)}^p(x,Q^2)$ which in turn, measure combinations 
of the parton longitudinal momentum distributions, $q_i(x,Q^2)$, $i=u,d,s,...$ at the scale $Q^2$.
% giving 
%the probability of finding a parton of type $i$ in the proton, with a fraction
%$x$ of its longitudinal momentum, at the scale $Q^2$.   
In the standard approach to DIS the $Q^2$ dependence of $F_2$ 
is described by the pQCD evolution equations 
\cite{DGLAP}, whose numerical solution requires parametrizing the input distributions 
at an initial scale $Q_o^2$ where pQCD is believed to be still applicable. 
%The values of the initial parameters, along with $\Lambda^2$ are  
%determined from global fits to the data. 
$Q_o^2$ serves as a boundary between the perturbative and non-perturbative domains,
although its value is somewhat arbitrary (in current parametrizations  
$Q_o^2 \approx 0.4 - 10$ GeV$^2$). 
We refer to this situation as the ``fixed initial scale'' description, and we
write explicitely the dependence of $q_i(x,Q^2,Q_o^2)$ on $Q_o^2$.
%% HYPOTHESIS OF FLUCTUATIONS
%%% SIMPLE KINEMATICAL ARGUMENT 
A simple kinematical argument shows that $Q_o^2$ is related to the invariant mass 
squared of the proton remnant after a parton is emitted, by: $M_X^2 \approx Q_o^2/x$.
%%
%% Write factorization property analogous to pomeron
%% 
At large $x$ 
%the mass of the intermediate system left behind is 
$M_X^2 \approx Q_o^2$ (at low $x$ 
this simple kinematical observation, and the factorization properties of the diffractive part of $F_2$,  
support the idea that the parton is emitted from a large mass object, 
indentified with a soft pomeron \cite{Wer}).   
We, therefore, explore the possibility 
that partons are not emitted directly from the nucleon, 
but that, before the pQCD radiative processes are initiated, a semi-hard phase occurs
where the dominant degrees of freedom are color neutral clusters with a mass distribution
peaked at: $\mu^2_{peak} \approx Q_o^2$.    
As a result, the nucleon structure function is related to the quark distribution  
by a smearing of the initial $Q_o^2$, namely    
%that $\mu^2$ undergoes fluctuations, namely:
%
\begin{equation}
F_2(x,Q^2) = x \sum_i e_i^2 \int_{\mu_o^2 > \Lambda^2}^{W^2} \frac{d \mu^2}{\mu^2} P(\mu^2) q_i(x,Q^2,\mu^2),
\label{cluster1}
\end{equation}
%%%%
where $P(\mu^2)$ ($P_{peak}(\mu^2) \approx P(Q_o^2))$ is the clusters' mass 
distribution, and the sum is extended
to valence quarks only since we are describing the large $x$ region. 
Eq.(\ref{cluster1}) expresses the fact that 
the initial stage of pQCD evolution is
characterized by color neutral clusters  
of variable mass, from which the hard scattering parton will emerge, in a subsequent stage
of the interaction.  

Our model is derived within the framework of the large $N_c$ approximation,
in analogy to the cluster hadronization schemes %first introduced 
implemented in the HERWIG Monte Carlo simulation \cite{HERWIG}.
In this scheme, hadronization proceeds as prescribed by the pQCD property of preconfinement
of color \cite{Ama}: at the end of the parton's pQCD evolution, color    
singlets are formed with a $Q^2$-- independent  mass (and spatial) distribution.  
%The large $N_c$ approximation supporting this mechanism generates 
%two basic assumptions of the model:  
%(1) the planar approximation; (2) the hypothesis 
%of no $q \bar{q}$ pair formation during pQCD evolution 
%\cite{Ama}. %or cluster formation.
In practical implementations \cite{HERWIG}, all gluons left at the hadronization 
scale, are ``forcibly'', or non-perturbatively, split into $q\bar{q}$ pairs.
It is this modification of the evolution equations that allows for 
the local parton-hadron conversion through preconfinement of colour: each color line 
``color--connects'' {\it e.g.} a quark to an anti-quark, forming  
a color singlet. The color-singlet clusters are then fragmented into hadrons.
In DIS the transiton hadrons $\rightarrow$ quarks $\rightarrow$ hadrons, is complicated
both by initial state radiation and by the presence of the beam cluster formed from the remnant 
of the initial hadron. This produces an additional rescattering term in the function $P$ in Eq.(2)
\cite{Liu_prep}. 
%Analogously to cluster formation in the final state, we assume that we          
%The conversion of a hadron into a parton through a cluster stage is described schematically 
%in Fig. 1(b).  

As a preliminary study, we considered both the low and the very large $W^2$ limits of Eq.(\ref{cluster1}). 
At low $W^2$, $F_2(W^2,Q^2) \approx P(Q^2)$, namely it is described by the behavior of the cluster 
distribution function. At very large $W^2$, $P_{peak}$ determines the value of the initial $Q_o^2$.
In Fig.1(b) we present our result for $F_2$ at $W^2 = 1.6$ GeV$^2$, where 
the cluster distribution was obtained from HERWIG \cite{HERWIG} (triangles), whereas the
red curve is an analytic calculation of the Sudakov-type behavior valid at large $Q^2$. 

\begin{figure}
\vskip -2.0cm
\hskip 2.0cm
\epsfig{figure=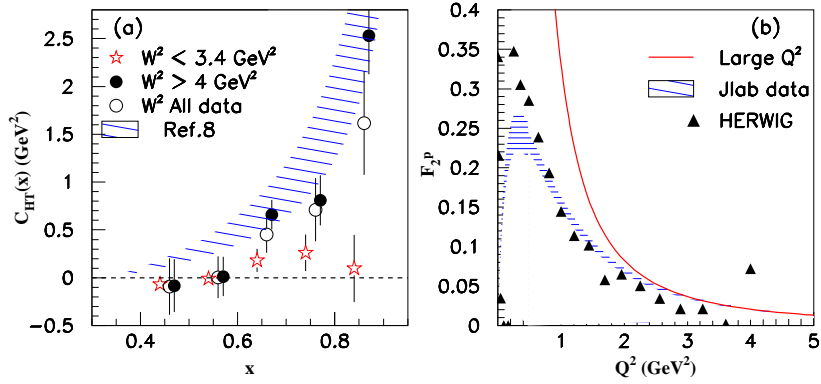,height=12cm}
\vskip -5.0cm
\caption{(a) Higher Twist coefficient from Eq.(2); (b) $Q^2$ dependence at 
fixed $W^2=1.6$ GeV$^2$.}
\end{figure}

\section*{Acknowledgments}
This work is supported by a research grant from the U.S. Department
of Energy under grant no. DE-FG02-01ER41200.

\end{document}